\documentstyle[12pt,aasms4,psfig]{article}
 
\slugcomment{to appear in Astrophysical Journal (letters)}
 
\begin{document}

\title{The Extreme Compact Starburst in MRK 273}
 
\author{C.L. Carilli and G.B. Taylor}
\affil{NRAO, P.O. Box O, Socorro, NM, 87801, USA \\}
\authoremail{ccarilli@nrao.edu}
 
\begin{abstract}

Images of neutral Hydrogen 21cm absorption and radio continuum
emission at 1.4 GHz from Mrk 273 were made using the Very Long
Baseline Array and Very Large Array. These images reveal a gas disk
associated with the northern nuclear region with a diameter of 0.5$''$
(370 pc), at an inclination angle of 53$^o$.  The radio continuum
emission is composed of a diffuse component plus a number of compact
sources.  This morphology resembles those of nearby, lower luminosity
starburst galaxies.  These images provide strong support for the
hypothesis that the luminosity of the northern source is dominated by
an extreme compact starburst.  The HI 21cm absorption shows an
east-west gradient in velocity of 450 km s$^{-1}$ across 0.3$''$ (220
pc), implying an enclosed mass of 2$\times$10$^{9}$ M$_\odot$,
comparable to the molecular gas mass.  The brightest of the compact
sources may indicate radio emission from an active nucleus (AGN), but
this source contributes only 3.8$\%$ to the total flux density of the
northern nuclear region.  The HI 21cm absorption toward the 
southeast radio
nucleus suggests infall  at 200 km s$^{-1}$ on scales
$\le$ 40 pc, and the southwest near IR  nucleus is not detected in high
resolution radio continuum images. 

\end{abstract}
 

\section {Introduction}

Luminous infrared galaxies are the most numerous sources
with luminosities $\ge$ 10$^{11}$ L$_\odot$ in the nearby 
universe (Sanders and Mirabel 1996). The bulk of
the luminosity from these sources is infrared emission from warm dust.
A critical question concerning
these sources is whether the dust is heated by an active nucleus, or 
a starburst? Recent studies using near IR spectroscopy suggest that
the dominant dust heating mechanism in most luminous infrared galaxies
(80$\%$) is star formation (Genzel et al. 1998), although  AGN heating
may become significant 
for the highest luminosity sources
($\ge$ 10$^{12.3}$ L$_\odot$;
Veilleux  et al. 1999). 
This question has taken on new significance 
due to the recent discovery of a population of luminous infrared galaxies
at high redshift seen in deep sub-millimeter and millimeter imaging
surveys.  If these high $z$ sources are starbursts, then they may
dominate the 
cosmic star formation rate at $z > 2$ (Smail, Ivison, and Blain 1997, 
Barger et al. 1998, Hughes et al. 1998, Blain et
al. 1999, Eales et al. 1999, Bertoldi et
al. 1999).

The most direct evidence to date of a dominant starburst in  
a luminous  infrared galaxy is the discovery of a population of radio
supernovae in the nuclear regions of Arp 220 by Smith et al. (1998)
using high resolution imaging at 1.4 GHz. 
Radio observations are unique 
in this regard, since they are unobscured by dust and
allow for imaging with mas resolution. We have begun a program of
imaging the radio continuum emission and HI 21cm absorption in
luminous infrared galaxies using the Very Long Baseline Array and 
the Very Large Array  at resolutions ranging from 1 to
100 mas. Results on the Seyfert 1 galaxy Mrk 231
have been presented in Carilli, Wrobel, and Ulvestad (1998), 
Taylor et al. (1999), and Ulvestad, Carilli, and Wrobel (1998). 
Those data revealed the presence of an AGN driven
radio-jet source on pc-scales at the center of 
a (possibly star forming)
gas disk with a diameter of a few hundred pc, 
with about half the radio continuum emission coming from
the disk. 

In this letter we present the results on Mrk 273 at z = 0.0377.
Mrk 273 has an infrared luminosity of $\rm L_{FIR} = 1.3
\times 10^{12} L_\odot$ (as defined in Condon 1992), where we assume 
H$_o$ = 75 km s$^{-1}$ Mpc$^{-1}$. The optical galaxy has been
classifed as a Seyfert 2, LINER, and both (Baan et al. 1998, 
Colina, Arribas, and Borne 1999, Goldader et al. 1995), and it has a disturbed
morphology on kpc-scales, with 
tidal tails indicating a merger event within the last 10$^{8}$ years
(Knapen et al. 1998).  Near IR spectroscopy reveals strong PAH
features indicative of a starburst, but also high ionization lines
indicative of an AGN (Genzel et al. 1998, Lutz et al. 1998).  The
X-ray emission also presents a mixed picture, with evidence for a
highly absorbed hard component,
but possible Fe L emission at 0.8 keV indicating cool (0.4 keV) gas
(Iwasawa 1999).  Broad absorption in the HI 21cm line and OH megamaser
emission have also been detected in Mrk 273  (Baan, Haschick, and 
Schmelz 1985, Schmelz, Baan, and Haschick 1988).

The nuclear regions in Mrk 273 on sub-arcsecond scales are complex,
with a double nucleus on a scale of 2$''$ seen in the near IR
(Knapen et al. 1998, Majewski et al. 1993, Armus et al. 1990), 
and in the radio continuum
(Ulvestad and Wilson 1984, Condon et al. 1991, Knapen et
al. 1998, Coles et al. 1999). The most peculiar aspect of 
Mrk 273 is that only one of the nuclei (the northern source) is seen
in both the radio continuum and the near IR. The southeast nucleus is
detected in the radio continuum, but is very faint in the near IR,
although there may be a faint blue `star cluster' at this position
(Scoville et al. 2000). The southwest nucleus is seen in the near IR,
but shows only very weak, extended radio continuum emission (Knapen et
al. 1998). High resolution near IR imaging with the HST shows that
both the north and southwest IR peaks 
are redder than the surrounding galaxy, and that
the northern nucleus is redder than the southwestern nucleus (Scoville et 
al. 2000). 

Imaging of CO emission from Mrk 273 shows a peak at the northern
nucleus, with faint extended emission on scales of a few arcseconds
(Downes and Solomon 1998). Downes and Solomon derive a molecular gas
mass of $1 \times 10^{9}$ M$_\odot$ for the northern nucleus, and find
that the CO is most likely in a disk with size $< 0.6''$. From these
data they conclude that the northern nucleus of Mrk 273 is an
extreme compact starburst, with an IR
luminosity of $6 \times 10^{11}$ L$_\odot$ emitted from a region $<$
400 pc in diameter. This conclusion is supported by the  
0.2$''$ resolution images of the HI 21cm absorption presented in Coles et
al. (1999),  which reveal a velocity 
gradient along the major axis of the northern nucleus.

In this letter we present high resolution
imaging (10 mas to 50 mas) of the HI 21cm absorption
and radio continuum emission from Mrk 273. These data
confirm the existence of a  rotating gas disk with a diameter of
350 pc, and reveal a population of compact sources,
possibly composed of luminous radio supernovae and/or nested radio
supernova remnants.

\section{Observations}

Observations of Mrk 273 were made on May 31 and June 6, 1999 
with the Very Long Baseline Array (VLBA), including the
phased Very Large Array (VLA) as an
element in the very long baseline array. The pass band was 
centered at the frequency of the neutral hydrogen 21cm line 
at a heliocentric redshift of: z =  0.0377, or cz = 11300 km
s$^{-1}$. The total bandwidth was 16 MHz, using two orthogonal
polarizations, 256 spectral channels, and 2 bit correlation.
The total on-source observing  time was 13.4 hrs.

Data reduction was performed using the Astronomical Image Processing
System (AIPS) and AIPS++.  Standard {\sl a priori} gain calibration
was performed using the measured gains and system temperatures of each
antenna.  The compact radio source J1337+550 was observed every 5 minutes,
and this source was used to determine the initial fringe rates and
delays.  The source 3C 345 was used to calibrate the
frequency dependent gains (band pass calibration). The source J1400+621
was used to check the absolute gain calibration.  
The results showed agreement of observed and
expected flux densities to within 3$\%$.

After application of the delay and rate solutions, and band pass
calibration, a continuum data set for Mrk 273 was generated by
averaging off-line channels. This continuum data set was then used for
the hybrid imaging process, which involves iterative imaging and
self-calibration of the antenna-based complex gains (Walker 1985). The
final iteration involved both phase and amplitude calibration with a 3
minute averaging time for phases and 15 minutes for amplitudes.  The
self-calibration solutions were applied to the spectral line data
set. The spectral line data were then analyzed at various spatial and
spectral resolutions by tapering the visibility data, and by smoothing
in frequency.  The continuum emission was subtracted from the spectral
line visibility data using UVLIN.  Images of the line and continuum
data were deconvolved using the Clark `CLEAN' algorithm as implemented
in IMAGR.  For the radio continuum images we also employed the
multi-resolution CLEAN algorithm as implemented in AIPS++ (Holdaway
and Cornwell 1999).  Results were consistent for all image
reconstruction algorithms, and we present the naturally weighted
Clark CLEAN continuum images in the analysis below.
The full resolution of the naturally weighted images is 
10 mas. We also present images at 50 mas resolution made using
a Gaussian taper of the visibilities. 

\section{Results and Analysis}

The 1.368 GHz continuum image of Mrk 273 at 50 mas resolution is displayed in
Figure 1. The image shows that the northern nucleus is 
extended, with a major axis of 0.5$''$ and a minor axis of
0.3$''$. The region shows two peaks separated by 0.11$''$. 
We designate the western peak N1 and the eastern peak N2.
These two peaks can also be seen in near IR images of Mrk 273
(Knapen et al. 1998). 
The total flux density from this region is 86$\pm$9 mJy.
The southeastern  source, which we designate SE, 
is also extended over about 0.3$''$, 
with a total flux density of 40$\pm$4 mJy. 

Figure 2 shows the 1.368 GHz
continuum images of the northern and southeastern 
nuclei of Mrk 273 at 10 mas resolution. The northern source
is highly resolved, consisting of a diffuse component extending over
0.5$''$, punctuated by a number of compact sources. Table 1 lists the
positions and surface brightnesses
at 10 mas resolution  of the six sources with surface brightnesses
$\ge$ 0.5 mJy beam$^{-1}$. 
Positions are relative to the peak surface brightness, corresponding
to N1. The nominal position of N1 in Figure 2 is (J2000):
$13^h 44^m 42.119^s$, $55^o 53' 13.48''$, based on 
phase-referencing  observations using the celestial calibrator 
J1337+550 with a 5 minute cycle time. Note that the minimum error in
the absolute astrometry is 12 mas, as set by the uncertainty in the 
calibrator source position (see Wilkinson et al. 1998 
and references therein). The true error after phase transfer is likely 
to be significantly higher than this (Fomalont 1995, Beasley and
Conway 1995).

Given the incomplete Fourier spacing 
coverage for VLBI imaging, in particular for short spacings, 
it is possible that the CLEAN algorithm has generated spurious
point sources when trying to deconvolve extended emission 
regions. Conversely, we cannot rule-out the possibility that
the extended emission is composed 
of mostly faint point sources. The use of multiresolution CLEAN
mitigates these problems, and the sources listed in Table 1
all reproduce with essentially the same surface brightnesses
for images made with the Clark CLEAN, multi-resolution CLEAN,
and for images made with different visibility weighting schemes.
The brightness temperatures of these sources
are all $\ge 3 \times 10^6$ K, indicating non-thermal emission. 
The southeastern nucleus is also resolved, with high surface brightness 
emission occuring over a scale of 50 mas. 
We set a 4$\sigma$ limit of 0.14 mJy
to any compact radio source associated with the southwestern
peak (large cross in Figure 1) seen at near IR wavelengths 
(Knapen et al. 1998).

Spectra of the HI 21cm absorption toward SE, and N1 and N2,
at 50 mas resolution
are shown in Figure 3. The spectrum of SE shows
a double peaked profile, with the two lines separated by 
400 km s$^{-1}$, each with a Full Width at Half Maximum (FWHM)
of about  280  km s$^{-1}$. 
There is marginal evidence that each component has
velocity sub-structure, but the SNR of these data are insufficient
to make a firm conclusion on this point. 
The peak optical depth of each line is about 0.12$\pm$0.02, and the 
implied HI column density in each component is then: N(HI) = 
6.4$\pm$1.1 $\times$10$^{19}$ $\times$ $T_s$ cm$^{-2}$, 
where $T_s$ is the HI spin temperature in K.

An interesting comparison is made with the MERLIN absorption spectra
at 0.2$''$ resolution toward the SE component (Coles et al. 1999).
At this resolution, MERLIN detects 19 mJy of continuum emission, and
shows a 3 mJy absorption line at about 11200 km s$^{-1}$, and weaker
absorption of about 1 mJy at 11400 km s$^{-1}$.  The VLBA data
show a peak continuum surface brightness of 10 mJy beam$^{-1}$ at 50 mas
resolution, and absorption line depths of 
1 mJy at both velocities.  This suggests that the
absorption at 11200 km s$^{-1}$ is due to extended gas covering 
both the compact and extended continuum emitting regions, while the 11400
km s$^{-1}$ absorption is due to a small cloud ($\le$ 40 pc) covering
only the high surface brightness continuum emission. Assuming 11200 km
s$^{-1}$ indicates the systemic velocity of the gas at that 
location in the galaxy disk (Coles et al. 1999),
then the higher velocity system would be infalling at 200 km
s$^{-1}$.

The spectrum of N2 shows a relatively narrow absorption line, with a
FWHM = 160 km s$^{-1}$, a peak optical depth of 0.59$\pm$0.06, and an
HI column density of $1.8\pm0.2 \times 10^{20}$ $\times$ $T_s$
cm$^{-2}$.  The spectrum of N1 shows a broad, flat absorption profile
with FWHM = 540 km s$^{-1}$, with optical depths ranging from
0.1 and 0.4$\pm$0.04 across the line profile.  
Again, there is marginal evidence for a few
narrower, higher optical depth components.  The total HI column density is
$1.8 \pm 0.3 \times 10^{20} \times T_s$ cm$^{-2}$. The velocity
range of the HI absorption toward N1 is comparable to that seen for
the OH megamaser emission (Baan, Haschick, and Schmeltz 1985,
Stavely-Smith et al. 1987).

Figure 4 shows the position-velocity (P-V) diagram for the HI 21cm
absorption along the major axis of the northern nucleus. There is a
velocity gradient from east to west of about 450 km s$^{-1}$ across
300 mas, plus an apparent flattening of the velocity distribution to
larger radii.  The P-V distribution is confused somewhat by the broad
absorption seen toward N1 (at position --90 mas in Figure 4). The
east-west velocity gradient of the HI absorption across the northern
source is consistent with results from  MERLIN HI 21cm 
imaging at 0.2$''$ resolution (Coles et al. 1999), and
with the velocity field derived from CO emission
observations at 0.6$''$ resolution  (Downes and Solomon 1998).

\section{Discussion}

The most significant result from our high resolution radio continuum
imaging of Mrk 273 is that the emission from the
northern nucleus extends over a region of $0.3'' \times 0.5''$
(220$\times$370 pc), punctuated by a number of compact sources with
flux densities between 0.5 and 3 mJy. This morphology resembles those
of the starburst nuclei of NGC 253 and M82 (Ulvestad and Antonnuci
1997, Muxlow et al. 1994), on a similar spatial scale. However, the
total radio luminosity is an order magnitude larger in Mrk 273.  The
physical conditions in this region are extreme, with a minimum
pressure of 10$^{-9}$ dynes cm$^{-2}$, and corresponding magnetic
fields of 100 $\mu$G.

The 1.4 GHz radio continuum emission from nuclear starburst galaxies
is thought to be primarily synchrotron radiation from relativistic
electrons spiraling in interstellar magnetic fields, with the
electrons being accelerated in supernova remnant shocks (Condon 1992,
Duric 1988). The compact sources are then individual supernovae or
supernova remnants, while the diffuse emission is thought to be from
electrons that have diffused away from the supernova remnant
shocks. Our high resolution images provide strong support for the
hypothesis of Downes and Solomon (1998) that the northern nucleus of
Mrk 273 is an extreme compact starburst, with a massive star formation
rate of 60 M$_\odot$ year$^{-1}$, as derived from the radio continuum
luminosity (Condon 1992), and occuring in a region of only 370 pc
diameter.  From their detailed analysis of the CO emission from Mrk
273, Downes and Solomon (1998) propose that the star formation occurs
in a disk with scale height of 21 pc and a total gas mass of $1 \times
10^9$ M$_\odot$.

The nature of the weak, compact radio continuum sources in Mrk 273 is
not clear, but given the similarity in morphology with the starburst
nuclei in M82 and NGC 253, it is likely that these sources are a
combination of nested supernova remnants and/or luminous radio
supernovae.  These sources have radio spectral luminosities $\ge
10^{28}$ ergs s$^{-1}$ Hz$^{-1}$ at 1.4 GHz, which is an order of
magnitude higher than the brightest radio supernovae remnants seen in
M82 (Muxlow et al. 1994), and are comparable in luminousity to the
rare class of extreme luminosity radio supernovae characterized by SNe
1986J (Rupen et al. 1987) and 1979C (Weiler and Sramek 1988).  A
substantial population of such luminous supernovae has been discovered
in the starburst nucleus of Arp 220 by Smith et al. (1998), who
suggest that the high luminosities of those supernovae may indicate a
denser local environment relative to typical supernovae, by a factor 3
or so (Chevalier 1984).  If the compact sources in Mrk 273 are nested
supernova remnants, then it would require 10 or more of the most
luminous M82-type supernova remnants in regions less than 7 pc in
size.  Future high resolution imaging of Mrk 273 is required to
clarify the nature of these compact sources.
 

It is possible that the brightest of the compact sources, coincident 
with N1, indicates the presence of a weak radio AGN. Supporting evidence
for this conclusion is the broad HI absorption line observed toward
N1. This  component contributes only 3.8$\%$ to the total 
radio luminosity at 1.4 GHz of the northern nuclear regions. 

From flattening of the radio spectrum between 1.6 and 5 GHz, Knapen et
al. (1998) suggested that there may be a dominant, synchrotron
self-absorbed radio-loud AGN in the northern nucleus of Mrk 273. The
images presented herein clearly preclude this hypothesis. We feel a
more likely explanation for the low frequency flattening is free-free
absorption. We are currently analyzing images with
sub-arcsecond resolution between 327 MHz and 22 GHz in order to
determine the origin of this low frequency flattening.

The gas disk hypothesis for the northern nucleus of Mrk 273 is
supported by the observed velocity gradient in the HI 21cm absorption
along the major axis.  The rotational velocity at a radius of 220 pc
is 280 km s$^{-1}$, assuming an inclination angle of 53$^o$. Assuming
Keplerian rotation, the enclosed mass inside this radius is then
2$\times$10$^{9}$ M$_\odot$, comparable to the molecular
gas mass observed on this scale.

Overall, these data support the idea that the dominant energy source
in the northern nuclear region
in Mrk 273 is a starburst and not an AGN.
However, the presence of an AGN somewhere in the inner 2$''$
of Mrk 273 is still suggested, based on 
the high ionization near IR lines (Genzel et al. 1998),
the (possible) hard X-ray component (Iwasawa 1999), and
the Seyfert II optical spectrum, although  Condon et al. (1991)
argue that a Seyfert II spectrum is not necessarily a conclusive AGN
indicator.  It is possible that the AGN
is located at either the SE radio nucleus, or the SW near IR
nucleus. 

The SE radio nucleus presents a number of peculiarities, the
most important of which is the weakness of the near IR emission
(Knapen et al. 1998, Scoville et al. 2000).  Knapen et al. (1998)
suggested that this source may simply be the chance projection of a
background radio source. However, the probability of a chance
projection of a 40 mJy source within 1$''$ of the northern nucleus is
only $4\times10^{-7}$ (Langston et al. 1990, Richards et al. 1999).
This low probability, and the fact that we see evidence for gas infall
into the SE nucleus in the HI 21cm absorption images, 
effectively preclude the background source
hypothesis.  The radio morphology is consistent with an amorphous jet,
or a very compact starburst, although the lack of CO emission from
this region argues for an AGN. 
One possible cause for the lack of
near IR emission is that the active region is still obscured at 2.2
$\mu$m.  The HI 21cm absorption column density is 
6.4$\pm$1.1 $\times$10$^{22}$ $\times$ ($\rm {T_s}\over{10^3 K}$) cm$^{-2}$,
while the absorption column derived from 
the hard X-ray spectrum may be as large as $4 \times
10^{23}$ cm$^{-2}$, depending on the X-ray powerlaw index.
Using the HI 21cm column leads to A$_v$ = 
$40 \times ({\rm {T_s}\over{10^3 K}})$, assuming a Galactic
dust-to-gas 
ratio. This is comparable the extinction responsible for the
obscuration in the near IR of the AGN in the powerful radio galaxy
Cygnus A (Ward 1996).  Imaging at wavelengths of 
10 $\mu$m or longer, with
sub-arcsecond resolution, is required to address this interesting
question.

We do not detect any high surface brightness radio emission associated
with  the SW near IR nucleus. This could simply mean that 
this region harbours a radio quiet AGN.
An alternative posibility is that  this is
a star forming region in which the 
star formation is very recent, commencing less than
10$^6$ years ago, such that a substantial population of radio 
supernovae and supernova remnants have  not yet had time to develop.

\vskip 0.2truein 

We thank J. Wrobel, J. Ulvestad, and  K. Menten for useful discussions
and comments. 
This research made use of the NASA/IPAC Extragalactic Data Base (NED)
which is operated by the Jet propulsion Lab, Caltech, under contract
with NASA. The VLA and VLBA are operated by the 
National Radio Astronomy Observatory, which  is a facility of
the National Science Foundation operated under cooperative 
agreement by Associated Universities, Inc. CLC acknowledges support from
the Alexander von Humboldt Society, and the Max Planck Institute for
Radio Astronomy.

\newpage

\centerline{\bf References}

Armus, L., Heckman, T.M., and Miley, G.K. 1990,
ApJ, 364, 471

Baan, W.A., Haschick, A.D., and Schmelz, J.T
1985, ApJ (letters), 298, 51

Baan, W.A., Salzer, J.J., and Lewinter, R.D. 1998, ApJ, 
509, 633

Beasley, A.J. and Conway, J.E. 1995, in {\sl Very Long Baseline
Interferometry}, eds. J. Zensus, P. Diamond, and P. Napier, p. 327

Bertoldi, F. et al. 1999, A\&A (letters), in preparation

Blain, A., Smail, I., Ivison, R.J., \& Kneib,
J.-P. 1999, MNRAS, 302, 623

Chevalier, R.A. 1984, ApJ (letters), 285, 63

Coles, G.H., Pedlar, A., Holloway, A.J., and Mundell, C.G. 1999,
MNRAS, 310, 1033

Colina, L., Arribas, S., and Borne, K.D. 1999, ApJ (letters),  527, 13

Condon, J.J. 1992, ARAA, 30, 575

Condon, J.J., Huang, Z.P., Yin, Q.F., and Thuan, T.X. 1991, ApJ 378, 65

Downes, D. and Solomon, P. 1998, ApJ, 507, 615

Duric, Neb 1988, Space Science Reviews, 48, 73

Iwasawa, K. 1999, MNRAS, 302, 961

Eales, S., et al. 1999, ApJ, 515, 518

Fomalont, E. 1995, in {\sl Very Long Baseline
Interferometry}, eds. J. Zensus, P. Diamond, and P. Napier, p. 363

Genzel, R. et al. 1998, ApJ, 498, 579

Goldader, J.D., Joseph, R.D., Doyon, R., and Sanders, D.B. 1995,
ApJ, 444, 97

Holdaway, M. and Cornwell, T. 1999, in preparation

Hughes, D. et al. 1998, Nature, 394, 341

Knapen, J.H. et al. 1998, ApJ (letters), 490, 29

Langston, G.I., Conner, S.R., Heflin, M.B., Lehar, J., and Burke,
B.F. 1990, ApJ, 353, 34

Lutz, D., Spoon, H.W., Rigopoulou, D., Moorwood, A.F., and Genzel, R. 1998,
ApJ (letters), 505, 103

Majewski, S.R., Hereld, M., Koo, D.C., Illingworth, G.D., 
and Heckman, T.M. 1993, ApJ, 402, 125

Muxlow, T.W., Pedlar, A., Wilkinson, P.N., Axon, D.J., Sanders,
E.M., and de Bruyn, A.G. 1994, MNRAS, 266, 455

Richards, E. 1999, ApJ, in press

Rupen, M.P., van Gorkom, J.H., Knapp, G.R., Gunn, J.E., and Schneider, D.P. 
1987, AJ, 94, 61

Sanders, D.B. and Mirabel, I.F. 1996, ARAA, 34, 749

Schmelz, J.T., Baan, W.A., and Haschick, A.D. 1988, ApJ, 329, 142

Scoville, N.Z. et al. 2000, AJ, in press (astroph 9912246)

Smail, I., Ivison, R., \& Blain, A. 1997, ApJ
(letters), 490, 5

Smith, H.E., Lonsdale, C.J., Lonsdale, C.J., and Diamond, P.J. 1998,
ApJ (letters), 493, 17

Stavely-Smith, L., Cohen, R.J., Chapman, J.M., Pointon, L., 
and Unger, S.W. 1987, MNRAS, 226, 689

Taylor, G.B., Silver, C.S., Ulvestad, J.S., and Carilli,
C.L. 1999, ApJ, 519, 185

Ulvestad, J.S., Wrobel, J.M., and  Carilli, C.L. 1999,
ApJ, 516, 127

Ulvestad, J.S. and Antonucci, R.J. 1997, ApJ, 488, 621

Ulvestad, J.S. and Wilson, A.S. 1984, ApJ, 278, 544

Veilleux, S., Sanders, D.B., and Kim, D.-C. 1999, ApJ, 522, 113

Walker, R.C. 1985, in {Aperture Synthesis in Radio Astronomy}, eds. 
R. Perley, F. Schwab, and A. Bridle (NRAO: Green Bank), p. 189

Ward, M.J. 1996, in {\sl Cygnus A}, eds. C. Carilli and D. Harris,
(Cambridge University Press), p. 43

Weiler, K.W. and Sramek, R.A. 1988, ARAA, 26, 295

Wilkinson, P.N., Browne, I.W.A., Patnaik, A.R., Wrobel, J.M., and
Sorathia, B. 1998, MNRAS, 300, 790

\newpage
\begin{deluxetable}{cc}
\footnotesize
\tablecaption{Compact Sources in Mrk 273}
\tablewidth{0pt}
\tablehead{
\colhead{Surface Brightness}  & \colhead{Relative Position}  \nl
\colhead{mJy beam$^{-1}$}  & \colhead{mas}  \nl
}
\startdata
3.07 &  0  ~~~  0  \nl
0.91 & 85E, 37N \nl
0.76 & 75E, 14N \nl
0.55 & 26E, 16S \nl
0.53 & 110E, 33N \nl
0.51 & 47W, 39S \nl
\enddata
\end{deluxetable}

\newpage

\centerline{Figure Captions}

\noindent Figure 1 -- An image of Mrk 273 at 1.368 GHz at 50 mas 
(37 pc) resolution.The contour levels are 
a geometric progression in the square root of
two, hence every two contours implies a factor two rise in 
surface brightness. The first contour level is 0.25 mJy beam$^{-1}$.
The peak surface brightness is 10 mJy beam$^{-1}$ and the off-source
rms is 85$\mu$Jy beam$^{-1}$.  
The reference position (0,0) corresponds to (J2000):
$13^h 44^m 42.142^s$, $55^o 53' 13.15''$, based on 
phase-referencing  observations using the celestial calibrator 
J1337+550 with a 5 minute cycle time. The cross in the SW corner
indicates the position of the SW near IR nucleus. 


\noindent Figure 2a -- An image of the northern nuclear regions of
Mrk 273 at 1.368 GHz at 10 mas (7.3 pc) resolution.
The contours are linear with an increment of 0.1  mJy beam$^{-1}$, 
starting at 0.1 mJy beam$^{-1}$.
The peak surface brightness is 3.05 mJy beam$^{-1}$
and the off-source rms is 36 $\mu$Jy  beam$^{-1}$. 

\noindent Figure 2b -- The same as Figure 2A, but now for the 
southeastern nuclear regions. The peak surface brightness is 
1.35 mJy beam$^{-1}$.

\noindent Figure 3 -- The HI 21cm absorption spectra toward Mrk 273
from images at 50 mas resolution. 
Figure 3a is the spectrum of N1. The peak surface brightness
of 7.8  mJy beam$^{-1}$ has been subtracted.
Figure 3b is the spectrum of  N2. The peak surface brightness
of 6.7  mJy beam$^{-1}$ has been subtracted.
Both these spectra have a velocity resolution of 29 km s$^{-1}$.
Figure 3c is the spectrum of SE at
a velocity resolution of 58 km s$^{-1}$.
The peak surface brightness
of 10 mJy beam$^{-1}$ has been subtracted. 
The zero point on the velocity scale corresponds to 
a heliocentric velocity of 11300 km s$^{-1}$ in all spectra.

\noindent Figure 4 -- The position-velocity plot for the HI 21cm absorption
across the major axis of the northern nucleus of Mrk 273 at a spatial
resolution of 50 mas (37 pc) and a velocity resolution of 60 km s$^{-1}$.  The
contour levels (in absoption) are: 0.4, 0.8, 1.2, 1.6, 2.0, 2.4, 2.8
mJy beam$^{-1}$.  The position of continuum component N1 corresponds
to --90 mas, while N2 corresponds to +20 mas.  The zero point on the
velocity scale corresponds to a heliocentric velocity of 11300 km
s$^{-1}$ in all spectra.

\vfill\eject

\begin{figure}
\psfig{figure=Fig1.ps,width=6in}
\caption{ }
\end{figure}

\vfill\eject       

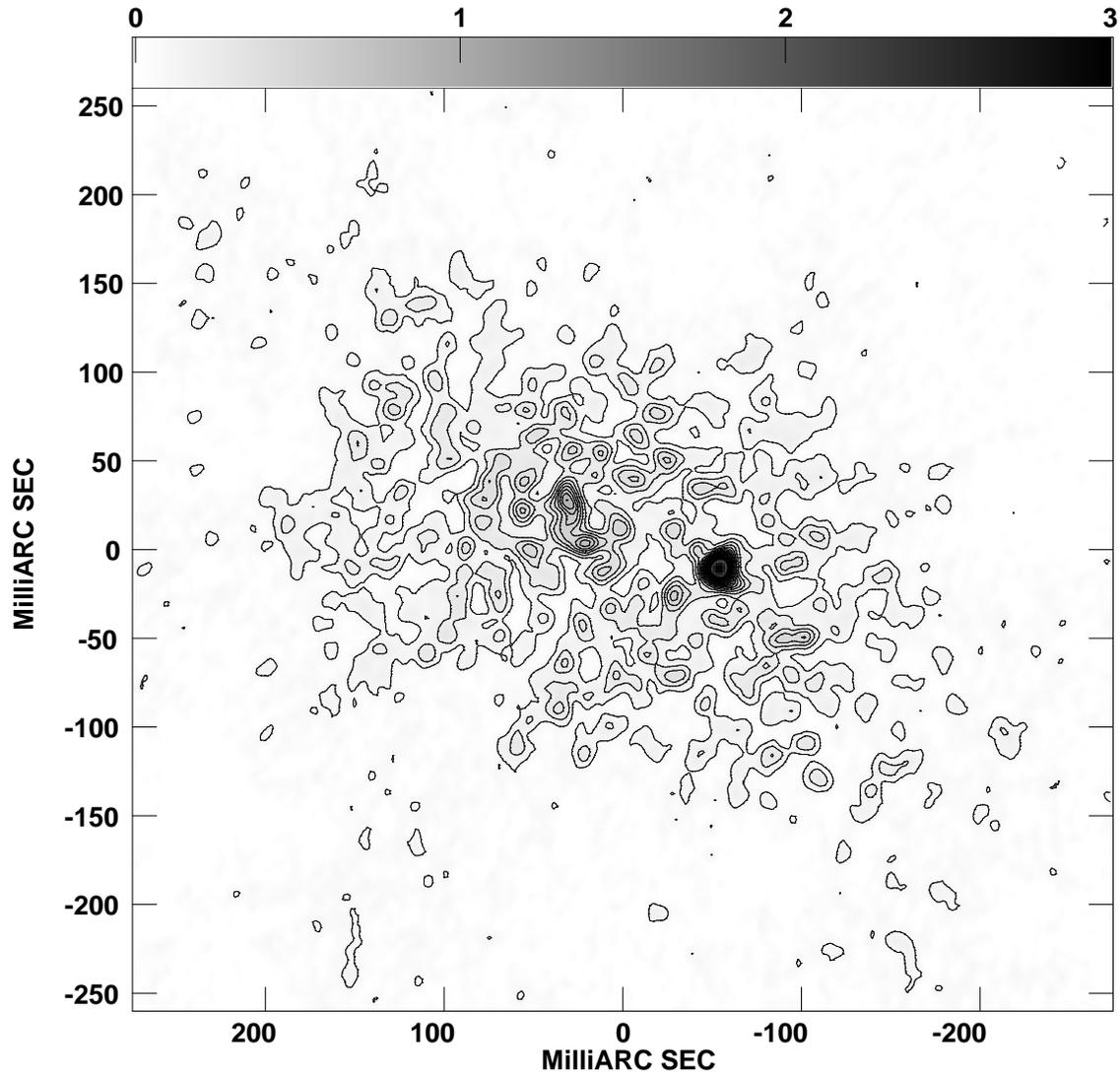
\begin{figure}
\psfig{figure=Fig2a.ps,width=6in}
\caption{ }
\end{figure}

\vfill\eject       

\begin{figure}
\psfig{figure=Fig2b.ps,width=6in}
\end{figure}

\vfill\eject       

\begin{figure}
\psfig{figure=N1.PS,width=3in}
\psfig{figure=N2.PS,width=3in}
\psfig{figure=S1.PS,width=3in}
\caption{ }
\end{figure}

\vfill\eject       

\begin{figure}
\psfig{figure=Fig4.ps,width=6in}
\caption{ }
\end{figure}

\vfill\eject       

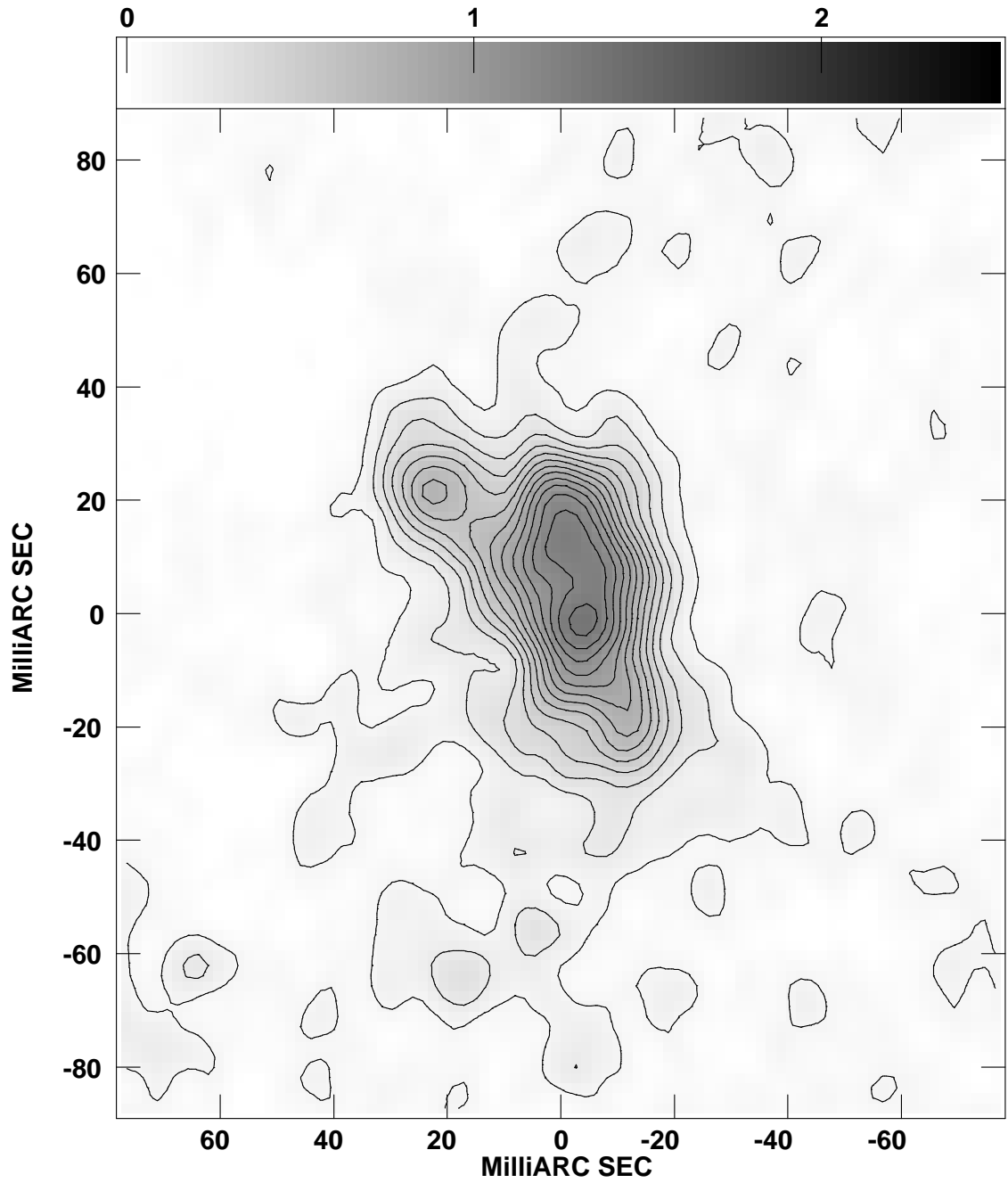
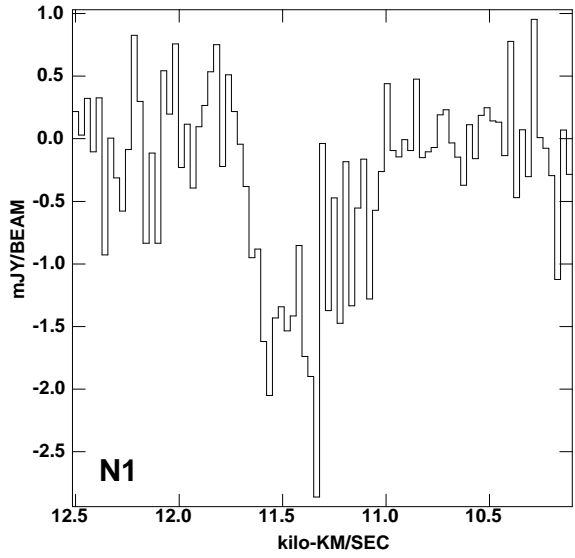
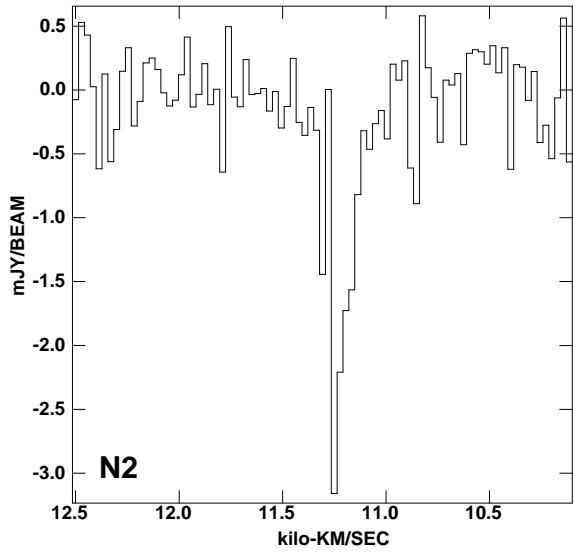
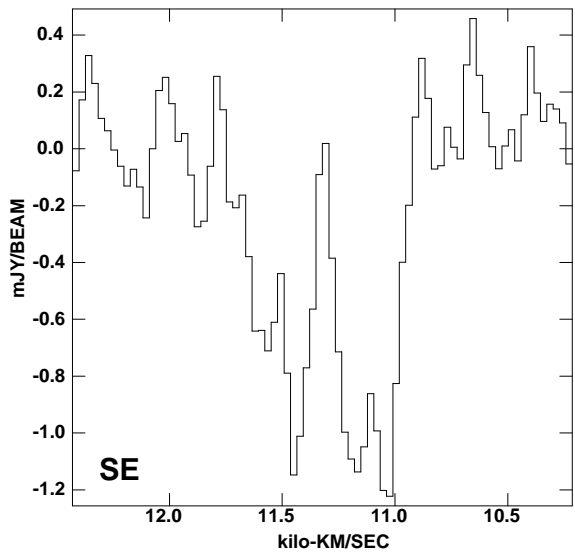
\end{document}